\documentclass[twocolumn,twoside,slac_two]{revtex4}
\usepackage{graphicx}
\usepackage{fancyhdr}
\newcommand{\lae}{\mathrel{<\kern-1.0em\lower0.9ex\hbox{$\sim$}}}
\newcommand{\gae}{\mathrel{>\kern-1.0em\lower0.9ex\hbox{$\sim$}}}

\begin{document}

%Title of paper
\title{Fermi detected blazars seen by INTEGRAL}

% Repeat the \author .. \affiliation  etc. as needed
%
% \affiliation command applies to all authors since the last
% \affiliation command. The \affiliation command should follow the
% other information

\author{Volker Beckmann}
\affiliation{Centre Fran\c{c}ois Arago, APC, CNRS/IN2P3, Universit\'e Paris Diderot, 10 rue Domon et Duquet, 75013 Paris}
\author{Claudio Ricci}
\affiliation{ISDC Data Centre for Astrophysics, 16 Ch. d'Ecogia, 1290 Versoix, Switzerland}
\author{Simona Soldi}
\affiliation{Laboratoire AIM - CNRS - CEA/DSM - Universit\'e Paris Diderot (UMR 7158), CEA Saclay, DSM/IRFU/SAp, 91191 Gif-sur-Yvette, France}

\begin{abstract}
Multiwavelength observations are essential to constrain physical parameters of the blazars observed by {\it Fermi}/LAT. 
Among the 187 AGN significantly detected in public {\it INTEGRAL} data above 20 keV, %by the imager IBIS/ISGRI, 
20 blazars were detected. 
15 of these sources allowed significant spectral extraction. 
They show hard X-ray spectra with an average photon index of $\langle \Gamma \rangle = 2.1 \pm 0.1$ and a hard X-ray luminosity of $\langle L_{20-100 \rm \, keV} \rangle = 1.3 \times 10^{46} \rm \, erg s^{-1}$. 15 of the {\it INTEGRAL} blazars are also visible in the first 16 months of the {\it Fermi}/LAT data, thus allowing to constrain the inverse Compton branch in these cases. Among others, we analyse the LAT data of four blazars which were not included in the {\it Fermi} LAT Bright AGN Sample based on the first 3 months of the mission: QSO~B0836+710, H~1426+428, RX~J1924.8--2914, and PKS~2149--306. 
Especially for blazars during bright outbursts, as already observed simultaneously by {\it INTEGRAL} and {\it Fermi} (e.g. 3C 454.3 and Mrk 421), {\it INTEGRAL} provides unique spectral coverage up to several hundred keV. We present the spectral analysis of {\it INTEGRAL} and {\it Fermi} data and demonstrate the potential of {\it INTEGRAL} observations of {\it Fermi} detected blazars in outburst by analysing the combined data set of the persistent radio galaxy Cen~A.

\end{abstract}

%\maketitle must follow title, authors, abstract
\maketitle

\thispagestyle{fancy}

% body of paper here - Use proper section commands
% References should be done using the \cite, \ref, and \label commands
% Put \label in argument of \section for cross-referencing
%\section{\label{}}

\section{INTRODUCTION}

The identification and modelling of {\it Fermi}/LAT detected blazars requires a dense observation of these sources accross the electromagnetic spectrum, from radio observations up to the TeV range. In the neighboring wavelengths of the LAT energy band blazars are well studied in the X-ray range up to 10 keV thanks to satellite based observations ranging from early missions like {\it EINSTEIN} and {\it EXOSAT} towards nowadays {\it Chandra} and {\it XMM-Newton}. The TeV range is now becoming more and more accessible through state-of-the-art projects like {\it HESS}, {\it MAGIC}, {\it VERITAS}, and {\it CTA} in the not too distant future. In order to determine the energy output of blazars in the inverse Compton branch, it is necessary though to fill also the parameter space below 100 MeV. At the time being, there is no mission available or even planned to access the energy range between 1 and 100 MeV, thus, in order to constrain the SEDs of blazars and to learn as much as possible about the low-energy end of the inverse Compton emission, we have to rely on the hard X-ray band. 
Blazars in the hard X-ray range (20 keV -- 500 keV) became accessible through instruments like {\it CGRO}/OSSE \cite{OSSEBlazar} and {\it BeppoSAX}/PDS \cite{BeppoBlazars} in the 1990s, and are nowadays probed mainly by experiments on-board {\it Swift} \cite{BATsurvey}, {\it Suzaku}, and {\it INTEGRAL} \cite{INTAGN2}. In this paper we focus on blazars commonly detected by {\it INTEGRAL} and {\it Fermi}/LAT and demonstrate the potential of {\it INTEGRAL} observations of blazars in outburst.  
The {\it INTEGRAL} and {\it Fermi}/LAT data of the IBIS/ISGRI detected blazars are described in Sect.~\ref{analysis}, followed by a discussion of the potential of {\it INTEGRAL} observations of blazars in outburst in Sect.~\ref{discussion}, and conclusions in Section~\ref{conclusions}.  

\section{{\it FERMI} AND {\it INTEGRAL} BLAZARS}
\label{analysis}

The {\it INTEGRAL} satellite \cite{INTEGRAL} hosts several instruments to study objects over a wide energy range. The spectrograph SPI performs high-resolution spectroscopy in the 20 -- 8000 keV range, and IBIS/ISGRI \cite{ISGRI} provides spectra and images from 18 keV up to several hundred keV. ISGRI is more sensitive and gives a higher spatial resolution than {\it Swift}/BAT, although the latter has a more uniform all-sky coverage due to its larger field of view and the {\it Swift} observation strategy following gamma-ray bursts. {\it INTEGRAL} provides broad-band coverage through the additional X-ray monitor JEM-X \cite{JEMX} in the 3--20 keV range and provides photometry with the optical camera OMC \cite{OMC} in the V-band. Since JEM-X and OMC have a much smaller field of view than ISGRI, for many sources no data are available from the monitors.
 
The analysis of the {\it INTEGRAL} data used here is described in detail in Beckmann et al. (2009). The second {\it INTEGRAL} AGN catalogue \cite{INTAGN2} lists in total 25 blazars which have been reported to be detected by IBIS/ISGRI above 20 keV, out of which 20 are detectable at $\gae 3 \sigma$ in the data analysed there. The results of a single power law model fit to the data taken between January 2003 and spring 2008 are given in Table~\ref{fitresults}. It has to be pointed out that the {\it INTEGRAL} data are not simultaneous with the {\it Fermi} data. In cases where the IBIS/ISGRI detection significance was $< 5 \sigma$, a photon index of $\Gamma = 2$ has been assumed in order to estimate the hard X-ray flux. For six of the 20 IBIS/ISGRI detected blazars, {\it INTEGRAL}'s X-ray monitor JEM-X gave a detection of $> 5 \sigma$ in the $3 - 20 \rm \, keV$ band: 1ES~0033+595, Mrk~421, 3C~273, H~1426+428, Mrk~501, and 3C~454.3. In addition, for six blazars V-band measurements are available, representing the average optical emission during the {\it INTEGRAL} observation.
\begin{table*}
\caption{{\it INTEGRAL} and {\it Fermi}/LAT average spectra. JEM-X and IBIS/ISGRI luminosities refer to the $3 - 20 \rm \, keV$ and $20 - 100 \rm \, keV$ energy band, respectively. Luminosities and fluxes are based on single power law model fits and are given in $[\rm erg \, s^{-1}]$ and $[\rm erg \, cm^{-2} \, s^{-1}]$, respectively. LAT detection significances and photon indices are based on the $0.1 - 200 \rm \, GeV$ energy band. The radio galaxy Cen~A is included for comparison.\\[0.5cm]}
\label{fitresults}
\begin{tabular}{lccccccccc}
%Name & $z$ &$\Gamma_{20 - 200 \rm keV}$ &  $\log L_{\rm JEM-X}$ & $\log L_{20 - 100 \rm \, keV}$ &  $\sigma_{> 100 \rm \, MeV}$ & $f_{0.1 - 100 \rm \, GeV}$ & $\log L_{0.1 - 100 \rm \, GeV}$ & $\Gamma_{> 100 \rm \, MeV}$ \\
Name & $z$ & $V [\rm mag]$ (OMC)& $\log L_{\rm JEM}$ & $\Gamma_{20 - 200 \rm keV}$  & $\log L_{\rm IBIS}$ &  $\sigma_{\rm LAT}$ & $f_{0.1 - 100 \rm \, GeV}$ & $\log L_{\rm LAT}$ & $\Gamma_{\rm LAT}$ \\
\hline
1ES 0033+595     & 0.086 &  & 45.08 & $3.6 {+0.4 \atop -0.3}$    & 44.36 & --      \\
IGR J03532--6829 & 0.087 &  &        & $3.5 {+0.7 \atop -0.6}$    & 44.63 & --      \\
PKS 0528+134     & 2.060 & $14.27\pm 0.09$&        &                            & 47.51 &  20 & $6.2 \times 10^{-11}$ & 48.68 & $2.85 \pm 0.06$ \\ 
QSO B0716+714    & 0.300 & &      &                            & 45.24 &  85 & $1.8 \times 10^{-10}$ & 46.90 & $2.11 \pm 0.02$ \\   
QSO B0836+710    & 2.172 & &      & $1.54 {+0.15 \atop -0.09}$ & 47.93 &  15 & $2.2 \times 10^{-11}$ & 48.42 & $3.1 \pm 0.1$ \\ % NEW   
Mrk 421          & 0.030 & $13.05 \pm 0.06$ & 44.50 & $2.45 {+0.03 \atop -0.02}$ & 44.92 & 107 & $3.3 \times 10^{-10}$ & 45.10 & $1.80 \pm 0.02$  \\
4C 04.42         & 0.965 & &       & $1.20 {+0.16 \atop -0.25}$ & 46.83 &  12 & $2.1 \times 10^{-11}$ & 47.23 & $2.6 \pm 0.1$    \\
3C 273           & 0.158 & $12.58 \pm 0.04$ & 45.90 & $1.92 \pm 0.03$            & 46.09 & 126 & $3.2 \times 10^{-10}$ & 46.62 & $2.76 \pm 0.02$  \\
PKS 1241--399    & 0.191 & &      &                            & 45.23 & --     \\
3C 279           & 0.536 & $15.5 \pm 0.1$      & & $1.6 \pm 0.2$              & 46.40 & 146 & $4.0 \times 10^{-10}$ & 47.85 & $2.36 \pm 0.01$  \\
H 1426+428       & 0.129 & $16.1 \pm 0.3$ & 45.48 &                            & 44.80 &  13 & $1.4 \times 10^{-11}$ & 45.01 & $1.65 \pm 0.14$  \\ % NEW
Mrk 501          & 0.034 & $13.21 \pm 0.04$ & 44.72 & $2.8 \pm 0.3$              & 44.05 &  49 & $1.0 \times 10^{-10}$ & 44.71 & $1.81 \pm 0.03$  \\ % MI
IGR J16562--3301 & 2.40 \cite{Masetti08}  &  &     & $1.32 {+0.16 \atop -0.07}$ & 47.80 & --               &\\
PKS 1830-211     & 2.507 &  &     & $1.49 {+0.05 \atop -0.07}$ & 48.19 &  53 & $2.2 \times 10^{-10}$ & 49.38 & $2.69 \pm 0.03$  \\
RX J1924.8--2914 & 0.352 &   &    &                            & 45.70 &  17 & $3.6 \times 10^{-11}$ & 46.39 & $2.37 \pm 0.06$  \\ % NEW
1ES 1959+650     & 0.048 &   &    & $1.9 \pm 0.4$              & 44.32 &  33 & $6.1 \times 10^{-11}$ & 44.79 & $2.12 \pm 0.05$  \\
PKS 2149--306    & 2.345 &   &    & $1.3 {+0.5 \atop -0.9}$    & 47.74 &  11 & $2.1 \times 10^{-11}$ & 48.48 & $3.1 \pm 0.1$    \\ % NEW
BL Lac           & 0.069 &   &    & $1.8 {+0.4 \atop -0.3}$    & 44.34 &  41 & $9.7 \times 10^{-11}$ & 45.32 & $2.45 \pm 0.03$    \\
IGR J22517+2217  & 3.668 &   &    & $1.4 \pm 0.4$              & 48.43 & --               \\
3C 454.3         & 0.859 & & 47.49 & $1.58 \pm 0.06$            & 47.76 & 232 & $7.9 \times 10^{-10}$ & 48.65 & $2.48 \pm 0.01$  \\
\hline
Cen A            & 0.002 & $13.5 \pm 0.5$ & 42.54 & $1.84 \pm 0.02$ & 42.70 & 34 & $8.9 \times 10^{-11}$& 42.09 & $2.88 \pm 0.05$ \\ 
\hline
%
%$^{a)}$ Masetti et al. 2008
\end{tabular}
\end{table*}

On average, the {\it INTEGRAL} IBIS/ISGRI detected blazars show the following properties. % in the {\it INTEGRAL} IBIS/ISGRI data. 
Redshift information is available for 18 of the sources, having an average luminosity of $\langle L_{20 - 100 \rm \, keV} \rangle = 1.3 \times 10^{46} \rm \, erg \, s^{-1}$, assuming isotropic emission. 15 sources allowed spectral fitting with $\langle \Gamma \rangle = 2.1$, and for 10 blazars the mass of the central black hole has been determined, giving $\langle M_{BH} \rangle = 6 \times 10^8 \rm \, M_\odot$. 

The analysis of {\it Fermi}/LAT data of the {\it INTEGRAL} detected blazars was done using Fermi Science Tools version 9.15.2 and applying the unbinned likelihood analysis. Data were used taken from the beginning of the {\it Fermi} mission until November 25, 2009, spanning about 16 months of LAT observations. We used the square root of the Test Statistic (TS) as an indicator for the significance of a source detection, and considered sources with $\sqrt{TS} > 10$ as detected.  Out of the 20 blazars listed in Tab.~\ref{fitresults}, 15 are detectable with a significance $> 10 \sigma$ in the LAT data analysed here. 11 of these LAT detected sources have already been reported in the {\it Fermi} LAT Bright AGN Sample (LBAS) based on the first 3 months of the mission \cite{FermiBSC,LBAS_SED}. The four blazars which were not included in that list are QSO~B0836+710, H~1426+428, RX~J1924.8--2914, and PKS~2149--306. 

\begin{figure}[t]
\centering
\includegraphics[angle=0,width=8cm]{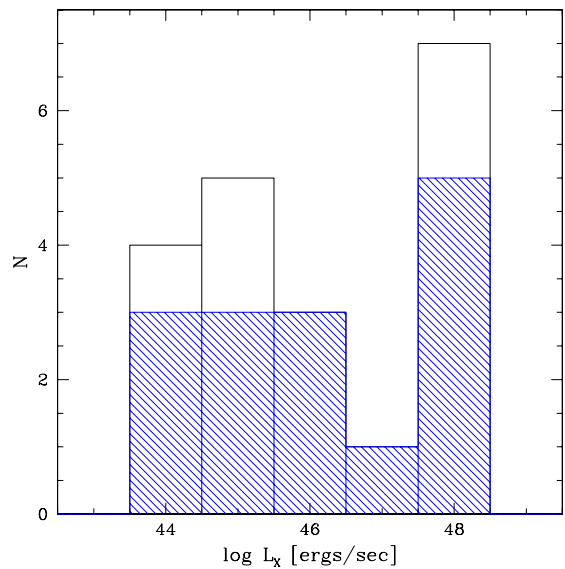}
\caption{Histogram of the hard X-ray ($20 - 100 \rm \, keV$) luminosity of the {\it INTEGRAL} detected blazars. The shaded area indicates those objects which are also detected by {\it Fermi}/LAT with $TS > 100$.} \label{lumihistogram}
\end{figure}

For all the 15 objects for which we have a significant LAT detection, the properties as reported in the {\it INTEGRAL} AGN catalogue are an average X-ray luminosity of $\langle L_X \rangle = 1.3 \times 10^{46} \rm \, erg \, s^{-1}$ and a photon index of $\langle \Gamma_X \rangle = 2.0$. Figure~\ref{lumihistogram} shows the histogram of hard X-ray luminosities, as derived from IBIS/ISGRI data, for the {\it INTEGRAL} detected blazars. The shaded area indicates those objects which are also detected by the LAT. 
In all 15 commonly detected blazars the gamma-ray luminosity in the $0.1 - 100 \rm \, GeV$ band is significantly larger than the hard X-ray ($20 - 100 \rm \, keV$) one, with an average ratio of $\langle L_\gamma / L_X \rangle = 6 {+4 \atop -2}$.  

\begin{figure}[t]
\centering
\includegraphics[angle=0,width=8cm]{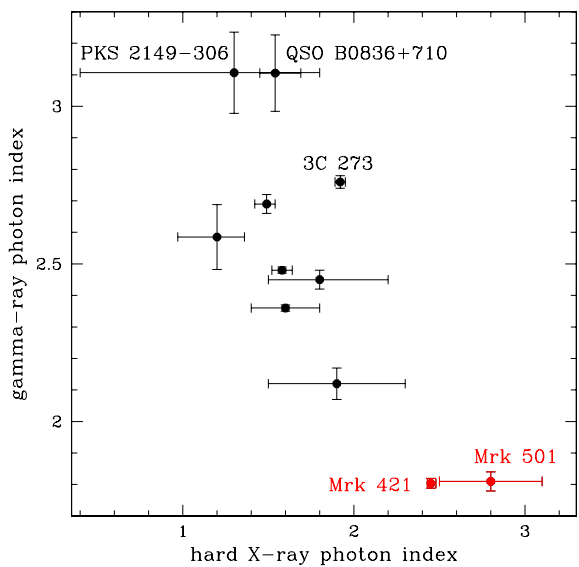}
\caption{Gamma-ray photon index (LAT) versus X-ray photon index (IBIS/ISGRI). Note that the data are not simultaneous.} \label{gammagamma}
\end{figure}
Mrk~421 and Mrk~501 are the only two blazars commonly detected which show a falling spectrum of the spectral energy distribution (e.g. $\log \nu f_\nu$ versus $\log \nu$ diagram) in the X-rays, but a rising one in the gamma-rays. All other objects show a rising X-ray ($\Gamma_X < 2$) and a falling gamma-ray spectrum ($\Gamma_\gamma > 2$). Although the data are not simultaneous, this indicates that in Mrk 501 and Mrk 421 we detect the synchrotron branch in the hard X-rays, while the gamma-rays are dominated by the inverse Compton component. In the other cases we see in both, the X- and the gamma-rays, the inverse Compton component with the peak of this component lying in between both bands ($200 \rm \, keV < E_{\rm IC \, peak} < 200 \rm \, MeV$). In these cases, the average difference between the photon indices is $\langle \Delta \Gamma \rangle = 0.9 \pm 0.3$, with the gamma-ray spectra being steeper. Obviously both, the X-ray and the gamma-ray photon index represent an average over many spectral states during which the photon index and flux can vary significantly -- see for example the case of the {\it Fermi}/LAT time resolved spectral analysis of 3C~273 \cite{3C273Fermi}. Figure~\ref{gammagamma} shows the comparison of the hard X-ray versus the gamma-ray photon index for a single power law fit. The results are consistent with the anti-correlation between X-ray and gamma-ray photon index as found for the {\it Fermi} LAT Bright AGN Sample (LBAS) \cite{LBAS_SED}, although the sample size we present here is not sufficient to confirm the trend. The two objects with the very steep ($\Gamma_\gamma > 3$) photon index in the gamma-rays are QSO~B0836+710 and PKS~2149--306, two of the fainter blazars which are not included in the LBAS. It can be expected that the ongoing {\it Fermi} mission will detect more of these soft gamma-ray blazars which belong to the LBL class as the survey sensitivity improves. 

\section{DISCUSSION}
\label{discussion}

\begin{figure}[t]
\centering
\includegraphics[angle=0,width=8cm]{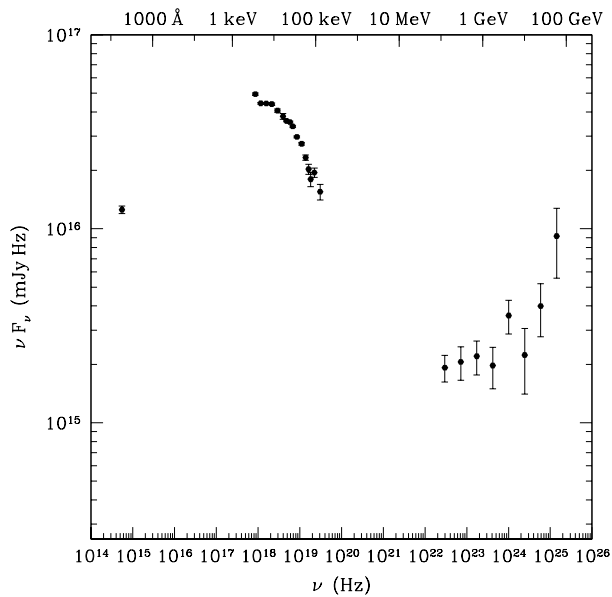}
\caption{Spectral energy distribution of Mrk 421 based on {\it INTEGRAL} OMC (optical V-band), JEM-X (3-30 keV), IBIS/ISGRI (20 -- 500 keV), and non-simultaneous {\it Fermi}/LAT data. The {\it INTEGRAL} data are dominated by measurements during the outburst in June 2006.} \label{Mrk421_SED}
\end{figure}

Although the {\it INTEGRAL} and {\it Fermi} data presented here have not been taken simultaneously, they allow already a rough outline of the spectral energy distribution of this blazar sample. Mrk 421 and Mrk 501 are mainly observed in outburst and they are both high-frequency peaked BL Lac objects, therefore the peak energy of the synchrotron and inverse Compton emission is at higher energies compared to the other blazars. We show as an example the spectral energy distribution of Mrk~421 based on {\it INTEGRAL} OMC, JEM-X, and IBIS/ISGRI data together with the {\it Fermi}/LAT 16-month average spectrum in Fig.~\ref{Mrk421_SED}. The {\it INTEGRAL} data shown here are dominated by an outburst observed in June 2006 \cite{Lichti08}, thus the X-ray emission level is about 10 times higher than in recent simultaneous X-ray and gamma-ray observations \cite{LBAS_SED}.
In Mrk~421 and Mrk~501 the (hard) 
X-rays are dominated by the synchrotron radiation, while the hard X-rays of the other objects and the gamma-rays of all 15 objects are dominated by inverse Compton emission.

The data presented here show also the potential for simultaneous {\it INTEGRAL} observations accompagnying an outburst in the gamma-rays of one of the blazars. Data in outburst often allow to determine the shape of the gamma-ray spectra within less than a week observing time. {\it INTEGRAL} observations of blazars in outburst have shown that significant spectra can be taken on time scales of $100 - 200 \rm \, ks$, for example in the case of 3C~454.3 \cite{3C454Pian} or Mrk~421 \cite{Lichti08}.

\begin{figure}[t]
\centering
\includegraphics[angle=270,width=8cm]{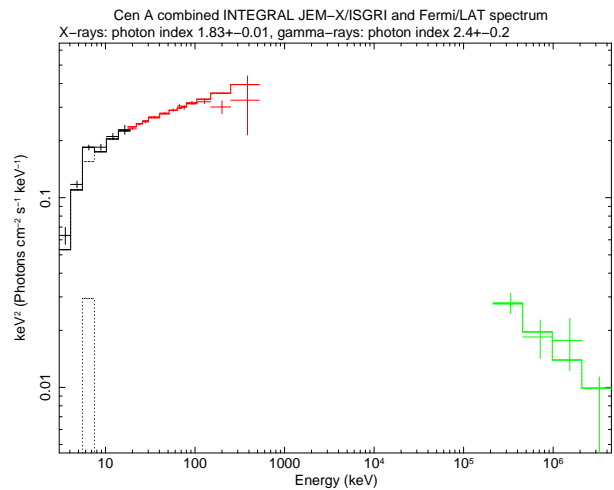}
\caption{The non-simultaneous data of Cen A can be combined using a simple broken power law model.} \label{CenAspectrum}
\end{figure}

\begin{figure}[t]
\centering
\includegraphics[angle=270,width=8cm]{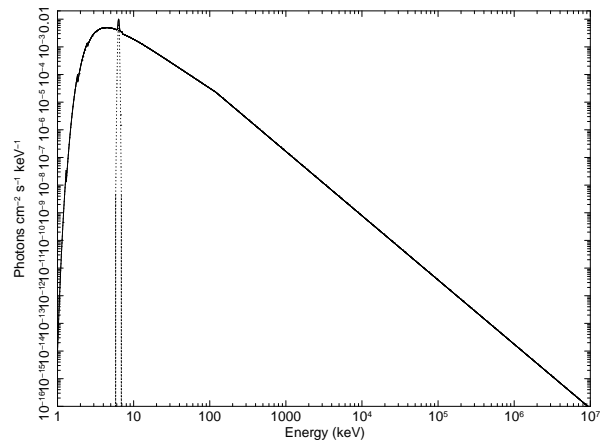}
\caption{The model sufficient to represent the X- and gamma-ray data of Cen~A connects both energy ranges smoothly.} \label{CenAmodel}
\end{figure}

As an example of what a simultaneous {\it INTEGRAL} and {\it Fermi} observation can provide, we studied the case of Cen~A. Although also the data for this low-redshift ($z = 0.0018$) radio galaxy are not taken simultaneously, the source shows low variability in the X-rays and gamma-rays, especially when compared to blazars. For the case of Cen~A, also {\it INTEGRAL}'s monitors JEM-X and OMC provide valuable data. The X-ray and gamma-ray data can be fit by a broken power law model absorbed by $N_H = 1.3 \times 10^{23} \rm \, cm^{-2}$, as shown in Fig.~\ref{CenAspectrum}. The photon index in the X-rays is $\Gamma_X = 1.83 \pm 0.01$, the one in the gamma-rays $\Gamma_\gamma = 2.4 \pm 0.2$ with a break energy at $E_{\rm break} = 135 {+161 \atop -27} \rm \, keV$. Note that the combined fit flattens the result for the gamma-ray photon index, which is $\Gamma_\gamma = 2.88$ when analysing the LAT data alone. The value of the break energy has to be taken with caution, as there are no data available between 500 keV and 200 MeV, and studies of the hard X-ray data do not show a significant cut-off \cite{Soldi05}. 
Nevertheless the combined modeling of the data as shown in Fig.~\ref{CenAmodel} gives a reliable estimate for the total energy output in the X-ray to gamma-ray domain of this elusive radio galaxy, with a total unabsorbed flux between 1 keV and 10 GeV of $f = 3.3 \times 10^{-9} \rm \, erg \, cm^{-2} \, s^{-1}$ and a luminosity of $L = 2.4 \times 10^{43} \rm \, erg \, s^{-1}$. In addition, the peak of the spectral energy distribution which in this model appears to be around a few hundred keV is consistent with the observations by {\it CGRO} OSSE, COMPTEL, and EGRET, which derived a peak energy around 500 keV for the intermediate emission state \cite{CGRO_CenA}. The peak flux is lower in our model, with $f_{\nu, \rm peak} \simeq 5 \times 10^{-7} \rm \, Jy$ compared to $f_{\nu, \rm peak} \simeq 8 \times 10^{-7} \rm \, Jy$ during the {\it CGRO} measurement.

\section{CONCLUSIONS}
\label{conclusions}

Out of the 20 blazars significantly detected within the {\it INTEGRAL} AGN catalogue, 75\% are already clearly detected by {\it Fermi}/LAT within the first 16 months of operation. Although this might appear to be a small fraction when compared to the total number of LAT detected blazars, it can be expected that gamma-ray blazars exhibiting a bright outburst will be detectable by {\it INTEGRAL} within 100 - 200 ks. This offers the possibility to constrain the spectral energy distribution between 3 keV and several hundred keV. The example of Cen~A, a radio galaxy which is quasi persistent in the X-ray and gamma-ray energy ranges, shows the potential of such a combined {\it INTEGRAL} OMC, JEM-X, and IBIS/ISGRI data set. With the on-going {\it Fermi} mission one will be able to detect even a larger fraction of hard X-ray blazars, especially those which show soft gamma-ray spectra such as QSO~B0836+710 and PKS~2149-306.

{\it INTEGRAL} is an ESA satellite open to observation proposals from astrophysicists from all countries. The next call for proposals\footnote{see http://www.sciops.esa.int/index.php?project=INTEGRAL} in 2010 with a dead line on April 23 (to be confirmed) offers again the possibility to apply for target of opportunity (ToO) observations of {\it Fermi} detected blazars in outburst. This way, {\it INTEGRAL} data can contribute significantly to the increasing number of well studied outbursts of blazars and help to understand this enigmatic type of AGN.

\bigskip % extra skip inserted
% Create the reference section using BibTeX:
%\bibliography{basename of .bib file}

\end{document}